\documentclass[smallextended]{svjour3}
            \usepackage{graphicx}
            \usepackage{mathptmx}
            \usepackage{latexsym}
            \journalname{quantum information processing}
            \begin{document}
            \title{Space and time dependent scaling of numbers in mathematical structures:  Effects on physical and geometric quantities.}
            \author{Paul Benioff}
            \institute{Physics Division, Argonne National
            Laboratory,\\ Argonne, IL 60439, USA \\
            Tel: 1-630-252-3218
            \email{pbenioff@anl.gov}}
            \date{Received: date / Accepted: date}
            \titlerunning{Number scaling effects on physics, geometry}
            \maketitle

            \begin{abstract}
             The relationship between the foundations of mathematics and physics is a topic of of much interest.  This paper continues this exploration by examination of the effect of space and time dependent number scaling on theoretical descriptions of some physical and geometric quantities. Fiber bundles provide a good framework to introduce a space and time or space time dependent number scaling  field.  The effect of the scaling field on  a few nonlocal physical  and geometric quantities is described. The effect on gauge theories is to introduce a new complex scalar  field into the derivatives appearing in Lagrangians.  U(1) invariance of Lagrangian terms does not affect the real  part of the scaling field.  For this field, any mass is possible. The scaling field  is also shown to affect quantum wave packets and path lengths, and geodesic equations even on flat space. Scalar fields described so far in physics, are possible candidates for the scaling field. The lack of direct evidence for the field in physics  restricts the scaling field in that the  gradient of the field  must be close to zero in a local region of cosmological space and time. There are no restrictions outside the region. It is also seen that the scaling field does not affect comparisons of computation or measurements outputs with one another.  However it does affect the assignment of  numerical values to the outputs of computations or measurements. These are needed because theory predictions are in terms of numerical values.
            \end{abstract}

            \section{Introduction}\label{Intro}

The relation between mathematics and physics at a foundational level is a topic of much interest. The need for deeper understanding of the relation is emphasized by Wigner's paper "On the unreasonable effectiveness of mathematics in the physical sciences" \cite{Wigner} This paper and others  \cite{Omnes,Plotnitsky} on this and similar topics \cite{Tegmark} suggest that this is a worthwhile area in which to work. Wigner's paper was also part of the impetus for earlier work towards a coherent theory of mathematics and physics together \cite{BenCTPM,BenCTPM2}.

It is clear that numbers play an essential role in this relation between physics and mathematics. Numbers provide the connection between theory and experiment.  Theoretical predictions of physical quantities as numbers  that are outcomes of computations, are to be compared with numerical outcomes of experiments.

This paper is based on the observation that numbers of different types, (natural numbers, integers, rationals, reals, and complex numbers), as mathematical structures satisfying relevant axioms, have a flexibility that affects both theory descriptions of physical quantities and the comparison of theory to experiment. This flexibility is based on the separation of two concepts, that of numbers as elements of a base set of a structure, and the values the numbers have in the structure.

A consequence of this concept separation  is the possibility  to assign different number values to each of the base set elements. This  is the basis of number scaling described here and in earlier work. This distinction between number and number value is exploited here by  first associating separate number structures  with each location in a space or space time manifold, $M$. This association allows the introduction of a space and time dependence of the number values associated with the elements of the base sets in the structures.

 The  separation of of number from number value is based on the  description of mathematical systems  of a given type as  structures or models.  Structures for each system type are required to satisfy a set of axioms relevant to the type of system \cite{Barwise,Keisler}.  Number structures with different number values for the same base set are referred to as scaled number structures or just scaled numbers.

 Another impetus for this work comes from gauge theories.  These are based on the association of separate finite dimensional vector spaces with each point of a space time manifold \cite{Montvay}. Gauge transformations between vector spaces at neighboring points of $M$ account for the freedom to choose basis sets for the spaces at different locations \cite{Yang}. Just one scalar field is assumed to be the common field for all the vector spaces.

   In this paper and in earlier work, this assumption is dropped in that a separate scalar field is associated with each point of $M$. The scalar fields are real for tangent vector spaces on $M$ and complex for vector spaces used in gauge theories. The scalar field correspondent to the basis choice freedom for vector spaces in gauge theories is based on the distinction between the concepts of number and number value, and the choice freedom for values of numbers in structures at different locations of $M$.

   The separation of the concepts of number from number value and the resulting number scaling are described in the next section.  This is followed in Section \ref{FB} by a description of fiber bundles \cite{Husemoller} on a base manifold, $M.$  Fiber bundles provide a very useful setup as a description of gauge theories \cite{Daniel,Drechsler} and as a distinction between local mathematical structures as fiber elements and global elements as fields or sections and connections on the bundle. The effects of number scaling are accounted for by use of a space and time dependent scaling field, $f$.  The effects of this field on gauge theories are described in Sections \ref{C} and \ref{GT}.

   Number scaling as defined here and in previous work \cite{BenNOVA,SPIE4,BenINTECH}, is different from other types of scaling described in the literature. It is different from conformal scaling \cite{Ginsparg,Gaberdiel} in that all  quantities, angles, as well as lengths, and other quantities are scaled. It is different from the geometric scaling introduced by Weyl \cite{Weyl} in that the scaling field introduced here is scalar and thus integrable.  A recent paper \cite{Czachor} describing the relativization of arithmetic is similar to the description  given here of number scaling for the natural numbers.

   The effects of number scaling  with fiber bundles are extended to other areas of physical theory in Section \ref{LNM} on local and nonlocal representations of nonlocal physical quantities. Nonlocal physical quantities are those expressed as integrals or derivatives over space and/or time. Local representations of these quantities are contained within a fiber as they are based on the structures in a fiber.  Nonlocal representations begin as integrals or derivatives over quantities in different fibers.  Since the integrands or derivative components belong to different fibers, connections are used to move the components to a common location. The expansion of a quantum wave packet  as an integral of space components and lengths of paths on $M$  are used as  examples.

   The space and time dependence of the scaling field might be expected to affect the comparison of computation and measurement results.  This is not the case.  This is shown in section \ref{CECR} where it is noted that comparison of results requires physical transmissions of the information in the outcome states to a common location for local comparison. It is also emphasized that the scaling field interpolates or connects experiment and computation outcomes as numbers in a base set with the theoretical values as number values in a scaled number structure.

   The next section discusses restrictions imposed on the scaling field, $f$ by the fact that experiments done to date do not seem to show the presence of $f$.  It is noted that $f$ should be roughly constant over a region of cosmological space and time in which sentient beings or observers can communicate.  Outside this region, which is small compared to the size of the universe,  there are no restrictions on $f.$

   Finally, last but by no means least, this paper is dedicated to Howard Brandt.  He encouraged this sort of work even though it did not seem to be relevant to  other areas in physics or mathematics.  His kind words will be missed.

            \section{Number scaling}\label{NS}

            As noted in the introduction, mathematics is based on a collection of structures of different types. Many of the structures are interconnected by maps between structures.  A structure  consists of a base set, a few basic operations, none, or a few basic relations and a few constants. The structure is required to satisfy a relevant set of axioms.  The different structure types are connected by  different maps between the structures. Maps such as scalar vector multiplication and vector norms, are examples of these maps between a vector space and the underlying scalar field structure.

            A good way to approach number scaling is to note that structures for the different types of numbers, as usually used, conflate two different concepts:  numbers as elements of the base sets, and the values that the numbers have in the structures.  As will be seen, one can define many different structures for the same type of number. The structures all have the same base set, but the elements of the base set have different values that depend on the structure containing them.

             Natural numbers, or nonnegative integers, provide the clearest example for the distinction between number and number value.   The usual structure,  $\bar{N}$ is represented by \begin{equation}\label{BN}\bar{N}=\{N,+,\times,<,0,1\}.\end{equation}Here $+,\times$ are basic operations, $<$  is a basic relation, and $0,1$ are constants. This structure satisfies the axioms for a commutative semiring with identity \cite{Kaye}. In this structure, the element, $"n",$ in the base set has value, $n$. In particular the base set element, $"1"$, has value $1$ because it satisfies the multiplicative identity axiom. Quotes are used to emphasize the fact that the base set elements, as represented here, are numbers and not number values.

            Here and from now on, number structures that satisfy relevant axioms are denoted by an overline, as in $\bar{N}$.   Elements, such as $N$ without an overline, denote base sets.

            The even numbers also give a valid structure for the natural numbers.   The corresponding structure is defined as \begin{equation}\label{BN2}\bar{N}^{2} =\{N_{2},+_{2},\times_{2},<_{2}, 0_{2},1_{2}\}. \end{equation} Here $N_{2}$ is the subset of $N$ consisting of the elements $"0","2","4",\cdots.$
            The value of the  number, $"2"$, satisfies the multiplicative identity axiom in this  structure. As a result it has value $1.$ The subscript $2$ on  $1$ and on other structure components identifies the structures containing the components.

            This is a simple example of the separation of the concept of number, as a base set element, from the value it has  in a structure.  It shows that the element, $"2"$ in the base set has value, $1_{2}$ in $\bar{N}^{2},$ and value $2=2_{1}$ in $\bar{N}$. In general the base set element, $"2n",$ has value $n$ in $\bar{N}^{2}$.  The only element whose value is unchanged is $"0".$

            A useful way to represent this is by means of value maps,  For example $val("n")=val_{1}("n")=n$ denotes the value of $"n"$ in $\bar{N}.$ Note that $\bar{N} =\bar{N}^{1}.$  However $val_{2}("2n")=n_{2}$ is the value of $"2n"$ in $\bar{N}^{2}.$  Here $n_{2}$ is the same value in $\bar{N}^{2}$ as $n$ is in $\bar{N}.$

            Another example  of natural numbers is provided by Von Neumann's description of the natural numbers as the finite ordinals in set theory.  In this representation, the base set $N$ consists of the elements, $\phi, \{\phi\}, \{\phi,\{\phi\}\},\cdots.$ Here $\phi$ denotes the empty set.  The base set $N_{2}$ is a subset of $N$ consisting of the sets, $\phi, \{\phi,\{\phi\}\},\cdots.$  The set $\{\phi,\{\phi\}\}$ has value $2$ in $\bar{N}$ and value $1_{2}$ in $\bar{N}^{2}.$

            These arguments can be extended to natural number structures whose base set consists of multiples of $n.$ This structure is represented by \begin{equation}\label{BNn}\bar{N}^{n}=\{N_{n},+_{n},\times_{n},<_{n},0_{n},1_{n}\}, \end{equation} The base set, $N_{n},$ consists of the elements, $"0","n","2n",\cdots.$ The element $"n"$ in $N_{n}$ has value, $1_{n}$ in $\bar{N}^{n}$.  If $n$ is even, $"n"$ has value $(n/2)1_{2} =(n/2)_{2}$ in $\bar{N}^{2}$ and value $n=n_{1}$ in $\bar{N}.$

            This can all be expressed using value maps.  One has $val_{n}("mn")=m_{n}$. Also \begin{equation} \label{val2mn}val_{2}("mn")=val_{2,n}(val_{n}("mn"))=\frac{n}{2}\times_{2}m_{2}.\end{equation}Here $m_{n}$ is the same number value in $\bar{N}^{n}$ as $m_{2}$ is in $\bar{N}^{2}.$  If $a$ is a number in $N_{n}$ then $$val_{n}(a)=\frac{1}{n}a_{n}$$ where $a_{n}=val_{n}(na)$ is the value of $"na"$ in $\bar{N}^{n}.$

            This shows that, by themselves, the  natural numbers in the base set have no  intrinsic values ($"0"$ excepted). Numbers acquire values only as base set elements in a valid structure.

            These relations between $\bar{N}^{2}$ and $\bar{N}^{n}$ show that the components of the structure, $\bar{N}^{n},$ can be mapped onto those of $\bar{N}^{2}.$  This map is   the identity map on the base set elements in that they are unchanged.  In essence this map represents the element values  and components in one structure in terms of those in another structure.

            The action of this map converts $\bar{N}^{n}$ to a new structure, $\bar{N}^{n}_{2}$ in which the number values of the numbers in $N_{n}$  are those of $\bar{N}^{2}$ and the basic operations and relation of $\bar{N}^{n}$ are expressed in terms of those of $\bar{N}^{2}.$  The resultant structure is given by \begin{equation}\label{BNn2}\bar{N}^{n}_{2}=\{N_{n},+_{2},\frac{2}{n}\times_{2}, <_{2},0_{2},\frac{n}{2}1_{2}\}.\end{equation}

            The scaling of the multiplication operation shown here is not arbitrary.  it is a consequence of the requirement that the map preserves the truth of the arithmetic axioms.  It is a straightforward exercise to show that $\bar{N}^{n}$ satisfies the axioms if and only if $\bar{N}^{2}$ does if and only if $\bar{N}^{n}_{2}$ does. In particular, note that the value $(n/2)_{2}\times_{2}1_{2}$ satisfies the multiplicative identity axiom in $\bar{N}^{n}_{2}$ if and only if $(2/2)_{2}=1_{2}$ satisfies the axiom in $\bar{N}^{2}:$ $$(\frac{n}{2})a_{2}(\frac{2}{n}\times_{2})(\frac{n}{2})1_{2}=(\frac{n}{2})a_{2} \leftrightarrow a_{2}\times_{2}1_{2}=a_{2}.$$

            This is the essence of the basic ideas behind number scaling applied to the natural numbers.  It shows that for any element, $l$ in  $N_{n},$ the value assigned to $l$ in $\bar{N}^{n}$  is $1/n$ times the value assigned to $l$ in $\bar{N}^{1}=\bar{N}.$ If $a$ is a factor of $n,$ then the value of $l$ in $\bar{N}^{n}$ is $a/n$ times the value of $l$ in $\bar{N}^{a}.$

            This description of number scaling can be easily extended to other types of numbers such as the integers, and  the rational, real, and complex  numbers. For any rational, real, or complex scaling factor, $s$ the corresponding scaled structures for  the rational, real and complex numbers are given respectively  by

            \begin{equation} \label{BRas}\overline{Ra}^{s}=\{Ra, \pm_{s}, \times_{s}, (-)_{s}^{-1_{s}}, <_{s},0_{s},1_{s}\}\end{equation} that satisfy the axioms for the smallest ordered field \cite{Rational}, \begin{equation}\label{BRs} \bar{R}^{s}=\{R,\pm_{s},\times_{s}, (-)_{s}^{-1_{s}},<_{s},0_{s},1_{s}\}\end{equation} that satisfy the axioms for a complete ordered field \cite{Real}, and  \begin{equation}\label{BCs}\bar{C}^{s}=\{C,\pm_{s},\times_{s}, (-)_{s}^{-1_{s}},(-)_{s}^{*_{s}},0_{s},1_{s}\}\end{equation} that satisfy axioms for an algebraically closed field of characteristic $0$ \cite{Complex}. Recall that overlined symbols, as in $\bar{R}^{s},$ denote structures, symbols without overline, such as $R$, denote a base set.

            Note that, unlike the case for the natural numbers (and integers), the base sets are the same for  all values of $s.$ This is a consequence of the fact that rational, real, and complex numbers are closed under the inverse operation.

            The standard or usual representation of these number types is for the scaling factor $s=1.$  However, as with the natural numbers, the emphasis here is on the relations between number structures with different scaling factors, not on the absolute values of the scaling factors.

            To see this, let $t$ and $s$ be pairs of rational, real, and complex scaling factors.
            For rational, real, and complex  numbers, the structures corresponding to  $\bar{N}^{n}_{2}$ are
            \begin{equation}\label{BRaRCts}\begin{array}{c}\overline{Ra}^{t}_{s}=\{Ra,\pm_{s},\frac{\mbox{$s$}}
            {\mbox{$t$}}\times_{s}, \frac{\mbox{$t$}}{\mbox{$s$}}(-)^{-1_{s}},<_{s},0_{s},\frac{\mbox{$t$}} {\mbox{$s$}}1_{s}\}\\\\\bar{R}^{t}_{s}=\{R,\pm_{s},\frac{\mbox{$s$}}
            {\mbox{$t$}}\times_{s}, \frac{\mbox{$t$}}{\mbox{$s$}}(-)^{-1_{s}},<_{s},0_{s},\frac{\mbox{$t$}} {\mbox{$s$}}1_{s}\}\\\\\bar{C}^{t}_{s}=\{C,\pm_{s}, \frac{\mbox{$s$}}{\mbox{$t$}}\times_{s}, \frac{\mbox{$t$}}{\mbox{$s$}}(-)^{-1_{s}},\frac{\mbox{$t$}} {\mbox{$s$}} (-)^{*_{s}},0_{s},\frac{\mbox{$t$}}{\mbox{$s$}}1_{s}\}\end{array}\end{equation} These structure representations hold only if $t$ and $s$ are both positive or both negative.  If not, then $<_{s}$ is replaced by $>_{s}$ in the structures for $\overline{Ra}^{t}_{s}$ and $\bar{R}^{t}_{s}.$

            To help in understanding the structures for complex numbers, complex conjugation is added as a basic  operation even though it is not necessary. Also $\bar{C}^{t}$ satisfies the complex number axioms if and only if $\bar{C}^{t}_{s}$ does.

            The complex number structure, $\bar{C}^{t}_{s}$ may seem strange because, $t/s$ can be complex when viewed from outside the structure.   The value maps are quite useful here to understand the relations among the structure elements. For example one has \begin{equation}\label{valstv}val_{s,t}((val_{t}(ta))^{*_{t}})= \frac{t}{s}(a_{t}^{*_{t}})_{s}=\frac{t}{s}a_{s}^{*_{s}}.\end{equation} Here $a_{s}^{*_{s}}$ is the same number value in $\bar{C}_{s}$ as $a_{t}^{*_{t}}$ is in $\bar{C}^{t}.$

            It will be quite useful to give generic representations of the structure for rational, real, and complex numbers.  Let the structures, $\bar{S}^{t}$ and $\bar{S}^{t}_{s}$ where
            \begin{equation}\label{BSts}\begin{array}{c}\bar{S}^{t}=\{S,Op_{t},Rel_{t},K_{t}\}\\ \bar{S}^{t}_{s}=\{S,Op^{t}_{s},Rel^{t}_{s},K^{t}_{s}\}\end{array}\end{equation} be generic structures for the different types of numbers: rational, real and complex that are shown in Eqs. \ref{BRas} -\ref{BCs} and \ref{BRaRCts}. $Op$ denotes the set of basic operations, $Rel$ the set of basic relations, if any, and $K$ the set of constants. $\bar{S}^{t}_{s}$ gives the components of $\bar{S}^{t}$ in terms of those of $\bar{S}^{s}.$ This is shown by  $Op^{t}_{s},Rel^{t}_{s}, K^{t}_{s}$.  These components include scaling factors where appropriate. For vector spaces based on complex scalars, $S=C$, for spaces based on real scalars, $S=R.$

             These generic structures can be used to define value maps for the  number structures in Eq. \ref{BRaRCts}.  Let $val_{t}(a)$ be the number value associated with an arbitrary  number $a$ in $S$ in the structure $\bar{S}^{t}$.  Then the definitions of Eq. \ref{BRaRCts} show that \begin{equation}\label{vsa}val_{s}(a)=val_{s,t}(val_{t}(a))=\frac{t}{s}val_{t}(a)_{s}.\end{equation} Here $val_{t}(a)_{s}$ is the same number value in $\bar{S}^{s}$ as $val_{t}(a)$ is in $\bar{S}^{t}.$

            It is important to note that, the observation that the numbers of the base sets do not have intrinsic values, does not mean that one can arbitrarily assign values to the numbers.  The requirement of preservation of axiom truths means that the freedom of value assignment is restricted to an arbitrary scaling of the values assigned to the base set elements and basic operations, where needed.  This is shown by Eq. \ref{vsa} in that the map, $val_{s,t}$ depends only on $s$ and $t$ and is independent of $a.$

            The effects of number scaling extend to vector spaces.  Let\begin{equation}\label{BV}\bar{V}=\{V,\pm, \cdot,|-|,v\} \end{equation}denote the usual structure for a normed vector space. Here $v$ denotes an arbitrary vector value and $|v|$ denotes the real valued norm of $v$ in $\bar{S}$.  The associated scalar structure is $\bar{S}.$ Also $\cdot$ denotes scalar vector value multiplication.

            For each scaling factor, $t,$ define $\bar{V}^{t}$ by\begin{equation}\label{BVt}V^{t}=\{V,\pm_{t},\cdot_{t}, |-|_{t},v_{t}\}.\end{equation}Here $v=v_{1}$ is the same vector value in $\bar{V}$ as $v_{t}$ is in $\bar{V}^{t}.$ However if $v$ is the vector in $V$ whose value is $v_{1}$ in $\bar{V}$, then $tv$ is the vector in $V$ whose value is $v_{t}$ in $\bar{V}^{t}.$   As was the case for scalars, the components of the vector space, $\bar{V}^{t}$, expressed in terms of those of $\bar{V}^{s},$ are given by
            \begin{equation}\label{BVts}\bar{V}^{t}_{s}=\{V,\pm_{s},\frac{t}{s}\cdot_{s},\frac{s}{t}|-|_{s}, \frac{s}{t}v_{s}\}.\end{equation}

            \section{Fiber bundles}\label{FB}
            The fact that values of  the  numbers in the base set $S$ can be scaled by any number value, $s,$ suggests that the freedom of basis choice in gauge theories can be extended to include freedom of choice of scaling factors for number values.  A very useful framework for describing this and for extension of number scaling freedom to other areas of physics is provided by fiber bundles.

            A fiber bundle \cite{Husemoller} consists of a triple, $E,p,M$  where $E$ is a total space, $p$ is a projection of $E$ onto $M$, and $M$ is a manifold.  For each $x$ in $M$, $p^{-1}(x)$ is the fiber at $x.$ For the purposes of  this work $M$ is  taken to be either a Euclidean space  with or without time, or space time of special relativity.  In this case a fiber bundle becomes a product bundle of the form $\{M\times F,p,M\}.$  Here $F$ is the fiber and $p^{-1}(x)=F_{x}$ is the fiber at $x.$

            Many choices for the contents of $F$ are possible.  For gauge theories \cite{Drechsler,Daniel} $F=\bar{V}$ a vector space.  The close association of scalars with vector spaces suggests the inclusion of scalars with the vector space or spaces. The resulting fiber bundle is
            \begin{equation}\label{MFSV}\mathcal{SV}=\{M,(\bar{S}\times\bar{V}),p,M\}\end{equation}The fiber at $x$ is given by \begin{equation}\label{FSVx}p^{-1}(x)=F_{x}=x\times \bar{S}\times\bar{V}= \bar{S}_{x}\times\bar{V}_{x}.\end{equation}.

            Fibers can be expanded to include representations of $M$.  This can be done by use of charts, $\phi,$ that are open set preserving maps of $M$ onto $\textbf{R}^{n}.$ Here $\textbf{R}^{n}$ is a chart representation of $M$ where $n$ is the dimension of $M$ with  coordinate values labelled with real numbers \cite{Tegmark}.  Also, here, $n=3$ or $4$. Since $M$ is flat, an atlas of charts is not needed, one chart over all of $M$ is valid in that $\phi(M)=\textbf{R}^{n}.$

            The resulting fiber bundle is\begin{equation}\label{MFSVR}\mathcal{SVR}=\{M\times(\bar{S} \times\bar{V}\times\textbf{R}^{n}),p,M\}\end{equation} with fiber at $x$ given by
            \begin{equation}\label{FSVRx}F_{x}=\bar{S}_{x}\times\bar{V}_{x}\times\textbf{R}^{n}_{x}.
            \end{equation}

            Fibers can be extended to include number scaling. Let $s$ be a scaling factor.  Define the bundle,
            \begin{equation}\label{MFSVRs}\mathcal{SVR}^{s}=\{M\times \bar{S}^{s}\times\bar{V}^{s} \times\textbf{R}^{n},p,M\}.\end{equation} Collect the fibers of these bundles for all $s$ into one fiber.  The resulting bundle is given by \begin{equation}\label{MFSVRcup}\mathcal{SVR}^{\cup}= \{M\times \bigcup_{s}(\bar{S}^{s}\times\bar{V}^{s})\times\textbf{R}^{n},p,M\}.\end{equation} The fiber at $x$ is given by \begin{equation}\label{FSVRxcup}F_{x}=\bigcup_{s}(\bar{S}^{s}_{x} \times\bar{V}^{s}_{x})\times\textbf{R}^{n}_{x}.\end{equation}

            One can define a structure group, $W_{S,V}$  on $\mathcal{SVR}^{\cup}.$ For each $t$ in $GL(1,S)$ the structure group element, $W_{S,V}(t),$ acts freely and transitively on $\bigcup_{s}(\bar{S}^{s}\times\bar{V}^{s})$ according to \begin{equation} \label{WSVs}W_{S,V}(t)(\bar{S}^{s}\times\bar{V}^{s})=\bar{S}^{ts}\times\bar{V}^{ts}.\end{equation}  This shows that \cite{Drechsler} $\mathcal{SVR}^{\cup}$ is a principal fiber bundle.  All scaling factors are equally likely.  There is no preference of one over another. Also \begin{equation} \label{WSVsu}W_{S,V}(s)\circ W_{S,V}(u)=W_{S,V}(su)=W_{S,V}(us)=W_{S,V}(u)\circ W_{S,V}(s). \end{equation}

            \section{Connections}\label{C}

            Fiber bundles of the form of $\mathcal{SVR}^{\cup}$ and further expansions of the fibers provide a suitable framework for the inclusion of the effects of space and/or time dependent number scaling into theoretical predictions of physical quantities. One way to include these effects is by use of a number scaling  or guide field $f$ with domain $M$ that takes values in $GL(1,S).$  This field can be represented by a pair of real scalar fields as in\begin{equation}\label{ftp} f(x)=e^{\theta(x)+i\phi(x)}.\end{equation}

            The effect of number scaling shows up in any physical quantity whose theoretical description is nonlocal.  These descriptions use  space and/or time integrals or derivatives. These quantities require mathematical combinations, such as addition or subtraction, of values in different fibers of a fiber bundle. Such combinations are not defined because mathematical operations on fiber elements are defined only within fibers.  They are not defined for elements in different fibers.

            This problem is remedied by the use of connections that connect elements in a fiber at one location to those at another location.  Connections are much used in gauge theories as elements of gauge groups that relate field values at neighboring points \cite{Montvay,Cheng}.  Here  number scaling connections are based on variations in $f$.

\begin{figure*}[h!]\begin{center}\vspace{1cm}
            \rotatebox{270}{\resizebox{200pt}{200pt}{\includegraphics[100pt,200pt]
            [440pt,550pt]{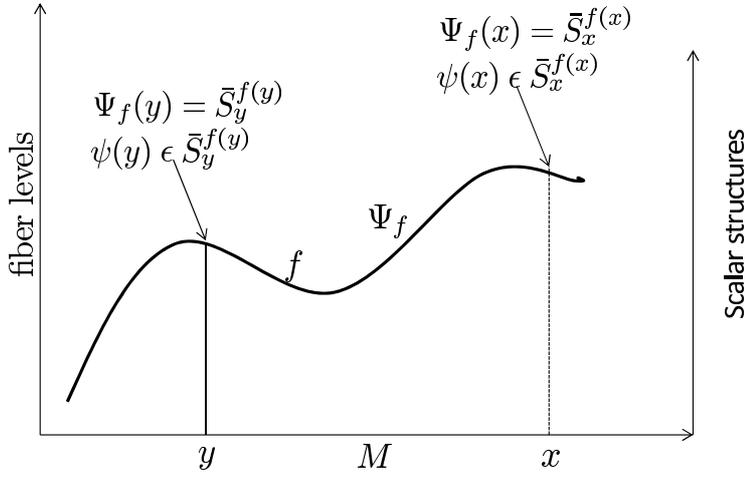}}}\end{center}
           \caption{Representation of the scaling field, $f,$ and values of the scalar structure valued field, $\Psi_{f}$ and the scalar field, $\psi,$ at two points, $x$ and $y$ of $M$.}\label{B1.eps}\end{figure*}

             An $f$ based connection affects both  scalar and vector structure valued fields as well as scalar and vector valued fields.  A schematic representation of the $f$ dependence for scalar structure and scalar valued fields is shown in Figure \ref{B1.eps}.

            Vector structure valued and vector valued fields are used to show the effect of $f$ on derivatives of these fields. Let $\Psi$ be a vector structure valued field over $M$.  In the presence of $f$, \begin{equation} \label{PsiV}\Psi(x)=\bar{V}^{f(x)}_{x}.\end{equation} The differential $d\Psi$ is represented by
            \begin{equation}\label{dPsi}d\Psi=\partial_{\mu,x}\Psi d^{\mu}x +\partial_{\mu,f,x}\Psi d^{\mu}x.
            \end{equation} Here \begin{equation}\label{pmuxPsi}\partial_{\mu,x}\Psi= \frac{\bar{V}^{f(x+d^{\mu}x)}_{x+d^{\mu}x}-\bar{V}^{f(x+d^{\mu}x)}_{x}}{d^{\mu}x}\end{equation} and \begin{equation}\label{pmufxPsi}\partial_{\mu,f,x}\Psi= \frac{\bar{V}^{f(x)+\delta f^{\mu}(x)}_{x}-\bar{V}^{f(x)}_{x}}{d^{\mu}x}.\end{equation}

            The derivative, $\partial_{\mu,x}\Psi$ is not defined because the terms in the numerator are in different fibers. The indicated subtraction  for structures is not defined between fibers at different locations.  It is defined between structures at the same location.

            This is remedied by the identity connection that maps $\bar{V}^{f(x+d^{\mu}x)}_{x+d^{\mu}x}$ into the same structure in the fiber at $x$.  Since this is the same structure as is $\bar{V}^{f(x+d^{\mu}x)}_{x},$ the indicated numerator subtraction gives $0$.  As a result, $\partial_{\mu,x}\Psi=0.$

             The term, $\partial_{\mu,f,x}\Psi,$ is not $0$ (unless $\partial_{\mu,x}f=0$) because the terms in the numerator of Eq. \ref{pmufxPsi} are in different levels of  the fiber at $x.$ This can be seen by parallel transporting $f(x+d^{\mu}x)$ to the same level value in the fiber at $x$ and defining the indicated level change, $\delta f^{\mu}(x)$ by \begin{equation}\label{fxdf}f(x)+\delta f^{\mu}(x)=f(x+d^{\mu}x). \end{equation}Taylor expansion of the right hand term  to first order in small quantities gives
             \begin{equation}\label{dfmux}\delta f^{\mu}(x)=\partial_{\mu,x}fd^{\mu}x. \end{equation}

             The derivative in Eq. \ref{pmufxPsi} is evaluated by  use of Eqs. \ref{fxdf} and \ref{dfmux} to  express the components of $\bar{V}^{f(x)+\delta f^{\mu}(x)}_{x}$ in terms of those in $\bar{V}^{f(x)}_{x}.$  The result is\begin{equation}\label{pfPsix}\partial_{\mu,f,x}\Psi=\frac{\bar{V}^{f(x)+\partial_{\mu,x}fd^{\mu}x}_{f(x),x}
             - \bar{V}^{f(x)}_{x}}{d^{\mu}x}=\frac{\bar{V}^{f(x)+\partial_{\mu,x}fd^{\mu}x-f(x)}_{f(x),x}}{d^{\mu}x} =\frac{\bar{V}^{\partial_{\mu,x}fd^{\mu}x}_{f(x),x}}{d^{\mu}x}.\end{equation}Here the equivalence $\bar{V}^{f(x)}_{x}=\bar{V}^{f(x)}_{f(x),x}$ has been used.

             The resulting scaling coefficient for $\bar{V}^{f(x)}_{x}$ is $\partial_{\mu,x}fd^{\mu}x/f(x).$ This is used to give the final result as \begin{equation}\label{pmfmuPsi}
            \partial_{\mu,f,x}\Psi=\frac{\partial_{u,x}f}{f(x)}\Psi=(\Gamma_{\mu}(x)+i\Delta_{\mu}(x))\Psi.
            \end{equation} Here \textbf{$\Gamma$} and \textbf{$\Delta$} are the gradients of the scalar fields, $\theta$ and $\phi.$  The same result holds if $\Psi$ is a scalar structure valued field where $\Psi(x)=\bar{S}^{f(x)}_{x}.$

            Scalar or vector valued fields, $\psi$ take values in scalar or vector structures at different points of $M$. At point $x$ of $M$,  $\psi(x)$ is either a vector or scalar in $\bar{V}^{f(x)}_{x}$ or in $\bar{S}^{f(x)}_{x}.$ Eq. \ref{dPsi} also holds for $\psi,$ in this case $\partial_{\mu,x}\psi$ is not $0.$  The other derivative remains the same.  The final result is that, in the presence of scaling,
            \begin{equation}\label{Dmuxp}D_{\mu,x}\psi=(\partial_{\mu,x}+\Gamma_{\mu}(x)+i\Delta_{\mu}(x))\psi.
            \end{equation}This result holds for both scalar and vector fields.

            \section{Gauge theories}\label{GT}

            Number scaling also has an effect on gauge theories. The freedom of basis choice at different points initiated by Yang Mills \cite{Yang} adds other factors to the connection used in the covariant derivative.

            For Abelian theories the covariant derivative of vector fields is given by Eq. \ref{Dmuxp} with another term added.  The result is \begin{equation}\label{DmuxpA}D_{\mu,x}\psi= (\partial_{\mu,x}+g_{r}\Gamma_{\mu}(x) +ig_{i}\Delta_{\mu}(x)+ih_{i} B_{\mu}(x))\psi.\end{equation} Here $\vec{B}$ denotes the photon field.  Coupling constants have been added to each of the component fields.

            The requirement that all terms appearing in Lagrangians be invariant under local $U(1)=e^{i\beta(x)}$ gauge transformations gives restrictions on the fields.  These are given  by \begin{equation}\label{grie}\begin{array}{l}
            g_{r}\Gamma'_{\mu}(x)+ig_{i}\Delta'_{\mu}(x)+ih_{i}B'_{\mu}(x)+i\partial_{\mu,x}\beta(x)\\\\\hspace{1cm}=g_{r}\Gamma_{\mu}(x)+ig_{i}\Delta_{\mu}(x)+
            ih_{i}B_{\mu}(x).\end{array}\end{equation}The fact that the Aharonov Bohm effect \cite{AhBo} requires that the photon field must be nonintegrable means that $\Delta_{\mu}(x)$ cannot be the photon field.  The reason is that it is integrable.

            Eq. \ref{grie} is satisfied by the conditions,\begin{equation}\label{GpG}\begin{array}{c} \Gamma'_{\mu}(x) =\Gamma_{\mu}(x)\\g_{i} \Delta'_{\mu}(x) +h_{i}B'_{\mu}(x)=g_{i}\Delta_{\mu}(x) +h_{i}B_{\mu}(x)-\partial_{\mu,x} \beta(x)\end{array}\end{equation}This equation can be satisfied in many ways.  The most general is to express $\beta(x)$ as the sum of two scalar functions as in $\beta(x)=\alpha(x)+\gamma(x)$ and split the second equation into two equations as in \begin{equation} \label{GpGa} \begin{array} {c}B'_{\mu}(x)=B_{\mu}(x)-\frac{\mbox{$1$}}{\mbox{$h_{i}$}} \partial_{\mu,x}\alpha(x) \\\Delta'_{\mu}(x)=\Delta_{\mu}(x)-\frac{\mbox{$1$}} {\mbox{$g_{i}$}} \partial_{\mu,x}\gamma(x). \end{array}\end{equation}

            The requirement that the photon field be massless is  taken care of here by setting $\gamma(x)$ to be nonzero and $x$ dependent.  There are no such restrictions on the $\Delta$ field.  If $\alpha(x)$ is $x$ dependent and nonzero, then $\Delta$ must be massless. Otherwise any mass, including $0,$ is possible for $\Delta.$  There are no mass restrictions on the $\Gamma$ field.

          The great accuracy of QED with no scaling fields present shows that the ratios of $g_{r}$ and $g_{i}$ to the fine structure constant must be very small. Alternatively, \textbf{$\Gamma$} and \textbf{$\Delta$} must be very small over regions of space and time in which QED is verified to be valid.

            It is of interest to speculate on the physical nature, if any, of the scalar fields, $\theta$ and $\phi.$ Could either or both be the Higgs field? \cite{Higgs}, the inflaton field? \cite{Inflaton}, dark matter? \cite{Dkmatter}, dark energy? \cite{Dkenergy},  some combination of these? \cite{Rinaldi,Bezrukov}, or none of these?  These are questions for the future.

            The appearance and effect of the \textbf{$\Gamma$} and \textbf{$\Delta$} fields in nonabelian theories is the same as that in Abelian theories.  Eq. \ref{grie} remains valid with additional requirements on the fields corresponding to the generators of the $su(n)$ Lie algebra.

            \section{Local and nonlocal expressions of physical quantities}\label{LNM}

            Fiber bundles offer an interesting way to distinguish between local and nonlocal theoretical descriptions of nonlocal physical quantities. This is a consequence of the observation that fibers of a bundle can be extended to include structures for  many different mathematical systems.  The setup with no scaling present is presented first to illustrate may of the properties of global and local representation.  This is followed by a discussion with scaling present.

            \subsection{No scaling present}

            Let the fiber of a bundle contain $\bar{C}\times\bar{H}\times \phi(M).$  Here $\phi(M)=\textbf{R}^{n}$ is a chart representation of $M$ in the fiber, $\bar{C}$ is a complex number structure, and $\bar{H}$ is a Hilbert space of vectors.
            The fiber bundle $\mathcal{CHM}$ is \begin{equation}\label{MFCHM}\mathcal{CHM}=M\times (\bar{C}\times\bar{H}\times\phi(M)),p,M.\end{equation} The bundle is a product bundle because $M$ is assumed to be flat as in Euclidean space (n=3) or in space time (n=4).  A fiber at point $x$ of $M$ is given by \begin{equation}\label{FxCHM}p^{-1}(x)=\bar{C}_{x}\times\bar{H}_{x}\times\phi_{x}(M). \end{equation}Here the  coordinate chart , $\phi(M)_{x}$ in the fiber at $x$ is taken to be the same as a direct chart map, $\phi_{x}$ of $M$ into the fiber at $x.$ That is, $\phi(M)_{x}=\phi_{x}(M)= \textbf{R}^{n}_{x}.$

            The fiber of the  bundle contains sufficient mathematical systems to represent a wave packet of a quantum system as a vector,  $\psi=\int\psi(y)|y\rangle dy$ in $\bar{H}.$ Here $\psi(y)$ is a complex amplitude or number value in $\bar{C}$ and the integral is over $\textbf{R}^{3}.$

            It follows from the bundle structure that fiber at each point $x$ of $M$ contains a representation of the wave packet as \begin{equation}\label{psixintx}\psi_{x}=\int_{x}\psi_{x}(y)|y\rangle_{x}dy.
            \end{equation}The subscript $x$ indicates that the integral is over $\textbf{R}^{3}_{x}$ and that the state  is a state vector in $\bar{H}_{x}.$

            This is defined to be a local representation of a nonlocal  physical quantity. A wave packet is a nonlocal quantity as it is a space integral of amplitudes and vectors. It is  local because the mathematical systems needed to describe it are in the bundle fiber.  Note that  the property of locality depends on the bundle fiber.  If the fiber did not contain $\bar{C}$ or $\bar{H}$ or $\phi(M)$ then  local representations of $\psi$ in the bundle would not be possible.  For this reason it may be better to relativize locality to a bundle by referring to the representation as a $\mathcal{CHM}$ local representation.

            A global representation of  the wave packet, as $\psi_{g},$ can also be given.   The goal is to lift the description of  $\psi_{g},$ as in integral over $M,$ up to  an integral in which each value of the integrand is in a different fiber of the bundle.  Connections are used to move the integrands to a common reference location in one fiber where the integral can be defined.

            The representation of $\psi_{g}$ as an integral over $M$ is given by $\psi_{g}=\int \psi_{g}(x)|x\rangle dx.$  Here $\psi_{g}$ can be regarded as a vector in a global Hilbert space, $H$, with $\psi_{g}(x)$ a complex number value in a global scalar field, $\bar{C}.$

             As before, let $\mathcal{CHM}$ be the bundle. The vector $\psi_{g}$ can be mapped into the bundle,  by first lifting each value, $\psi_{g}(x)|x\rangle$ of the integrand to a corresponding integrand,  $\psi_{g}(z)_{x}|z\rangle_{x},$ in $\bar{H}_{x}$ in the fiber of $\mathcal{CHM}$ at $x.$ The point, $z,$ in $\textbf{R}^{3}_{x}$ is related to the point, $x,$ in $M$ by $z=\phi(x)_{x}=\phi_{x}(x).$ Here $\psi_{g}(z)_{x}$ is the same complex amplitude (number value) in $\bar{C}_{x}$ as $\psi_{g}(x)$ is in $\bar{C}.$

\begin{figure*}[h!]\begin{center}\vspace{3cm}
            \rotatebox{270}{\resizebox{170pt}{170pt}{\includegraphics[190pt,200pt]
            [540pt,550pt]{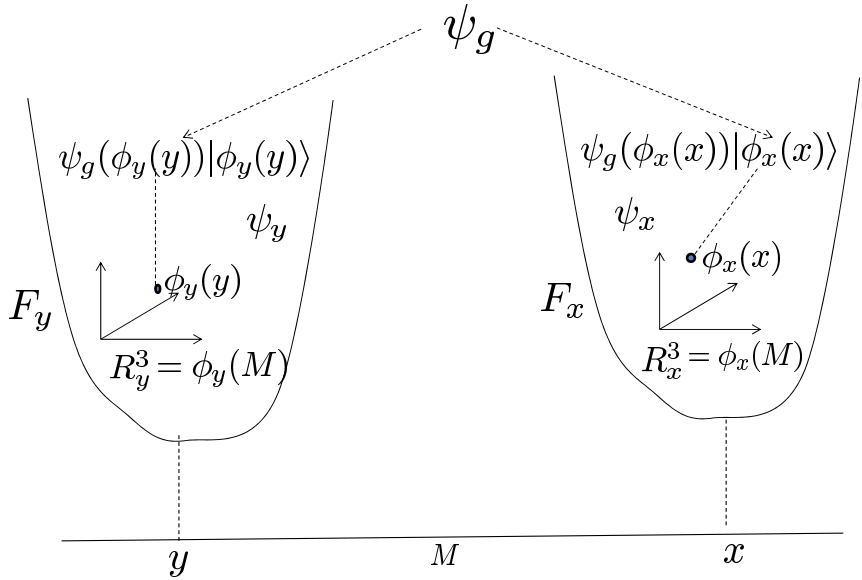}}}\end{center}
           \caption{Representation of  the global representation, $\psi_{g}$  with components in each fiber. Two components, one in the fiber at $y$ and the other in the fiber at $x$ are shown  Included also are the chart representations, $\phi_{y}(M),$ and $\phi_{x}(M)$, of $M$ in each of the two fibers.  Local representations of the wave packet, as integrals over $\phi_{x}(M)$ and $\phi_{y}(M),$  are indicated by $\psi_{y}$ and $\psi_{x}.$ }\label{B2.eps}\end{figure*}

            Figure \ref{B2.eps} shows schematically two components of the global representation of $\psi_{g}$ in fibers at points, $y$ and $x.$  Each fiber includes a chart representation $M$.  The local representations as integrals over the chart space are indicated by $\psi_{y}$ and $\psi_{x}.$

            An integral of  state components in the different fibers would not make sense because the indicated addition of integrands over different fibers is not defined. Addition is defined only for components within a fiber, not between components in different fibers. This problem is solved by the use of connections to map the integrands in the different fibers to integrands in one fiber at some reference location, $x_{0},$ on $M$.

            In the absence of scaling the connection is the identity map.  The action of this map on the component, $\psi_{g}(z)|z\rangle$ in the fiber at $x$ where $z=\phi_{x}(x),$  gives the component, $\psi_{g}(y)_{x_{0}}|y\rangle_{x_{0}}$ in the fiber at $x_{0}$ where $y=\phi_{x_{0}}(x).$

            The definition of the localized wave packet as an integral over $\phi_{x_{0}}(M)$ requires that the family of  charts $\phi_{x}$ for all $x$ in $M$ satisfy some consistency conditions.  For each pair of points $x,y,$ in $M$ one  requires that $\phi_{x}$ be the same chart at $x$ as $\phi_{y}$ is at $y$.  In particular for each point $z$ in $M,$ one requires that $\phi_{x}(z)$ be the same point in $\textbf{R}^{3}_{x}$ as $\phi_{y}(z)$ is in $\textbf{R}^{3}_{y}.$

            If these chart conditions are satisfied, one can define the $x_{0}$ localized representation of $\psi_{g}$ as \begin{equation}\label{psigx0w}\psi_{g}=\int_{x_{0}}\psi_{g}(w)|w\rangle_{x_{0}}dw.
            \end{equation}Comparison of this localized representation of $\psi_{g}$ with the local one shown in Eq. \ref{psixintx} with $x=x_{0}$ shows that $\psi_{g}=\psi_{x_{0}}.$  The localized representation of the global representation of the state is the same as the original local representation of the state.

            This example might lead one to think that there is no difference between global and local representations of nonlocal physical quantities.   As a result it makes no difference which representation one uses.  However this is the case with no number scaling present.  If number scaling is present then the localized version of the global representation is different from the local representation.
            \subsubsection{Global representations at different fiber levels}

             Global representations of the wave packet at different fiber levels requires extension of the fiber bundle to \begin{equation}\label{MFCHMC}\mathcal{CHM}^{\cup}=M\times \bigcup_{c}(\bar{C}^{c}\times\bar{H}^{c}\times \phi^{c}(M)),p,M.\end{equation} Here $\phi^{c}_{x}(M)=(\bar{R}^{c}_{x})^{3}.$ The fiber at each point, $x$ of $M$ is defined by \begin{equation}\label{MFCHMCp1x}p^{-1}(x)=\bigcup_{c} (\bar{C}_{x}^{c}\times\bar{H}_{x}^{c}\times \phi^{c}_{x}(M)).\end{equation}

            There are many different equivalent local representations of the wave packet, one at each level, $c,$ of the fiber. It is expressed by \begin{equation}\label{psixc}\psi_{x,c}=\int_{z}\psi_{x,c}(z)dz_{x}.\end{equation} The integral is over all $z$ in $\phi_{x}(M)$ and $\psi_{x,c}(z)$ is a number value in $\bar{C}^{c}_{x}.$

            The wave packets $\psi_{x,c}$ for different $c$ are all equivalent.  To see this let $W_{CH}$ be the structure group on $\mathcal{CHM}^{\cup}.$ For each complex number, $d$, in $GL(1,C)$ the action of the group element $W_{CH,d}$ is given by \begin{equation}\label{WCHd} W_{CH,d}(\bar{C}^{c}\times \bar{H}^{c}) =\bar{C}^{dc}\times\bar{H}^{dc}.\end{equation}It follows that $\psi_{x,dc}=W_{CH,d}(\psi_{x,c})$ is the same vector in $\bar{H}^{dc}_{x}$ as $\psi_{x,c}$ is in $\bar{H}^{c}_{x}.$

            The components of the  corresponding global representation, $\psi_{g,c}$ of the wave packet are similar to those shown in Fig. \ref{B2.eps}. The component $\psi_{g,c}(\phi_{y}(y))|\phi_{y}(y)\rangle_{c}$  in the fiber at $y$ is a vector in $\bar{H}^{c}_{y}$ and $\psi_{g,c}(\phi_{y}(y))$ is a complex number value in $\bar{C}^{c}_{y}.$ Connections are used to transform the integrands, all at level, $c$, to a fiber at a arbitrary reference location, $x_{0}$.  This is needed  so that the integral expression of $\psi_{g,c}$ makes sense.

            \subsection{Number scaling present}

            To account for scaling, one follows the same prescription as was used to obtain a representation of $\psi_{g}$ at a reference location, $x_{0},$ as in Eq. \ref{psigx0w}. The connection  to transform the wave function components in the fibers at each point $z$ of $M$ to $x_{0}$ uses  the scaling function $f$ shown in Fig. \ref{B1.eps}.  The result of the parallel transform multiplies the integrands by a factor $(cf(z))/cf(x_{0})$.  The resulting expression for the wave packet at $x_{0}$ is given by\begin{equation}\label{psicgx}\psi_{c,g,x_{0}}=\int_{x_{0}} \frac{cf(z)}{cf(x_{0})}
            \psi_{x_{0},c}(w) |w\rangle dw.\end{equation}  The integral is over all points, $w$, in $\phi^{c}_{x_{0}}(M)=(\bar{R}^{c}_{x_{0}})^{3}.$ Also $z=(\phi^{c}_{x_{0}})^{-1}(w).$

            This result shows that the global representation of the scaled wave packet is independent of the level $c.$  The expression is invariant under level change.  This is quite satisfactory since physical predictions and expressions of physical quantities as number values or vector values should not depend on the fiber level chosen for expression.

            An equivalent expression of $\psi_{g}$ ($c$ is suppressed) replaces $f$ by its scalar field representation as in Eq. \ref{ftp}. the result is \begin{equation}\label{psigx}\psi_{x_{0},g}=e^{-\theta(x_{0})-i\phi(x_{0})}\int_{x_{0}}e^{ \theta(w)+i\phi(w)}\psi_{x_{0},c}(w) |w\rangle dw.\end{equation}  Here the domain of $f(z)$ has been lifted from $M$ to $\phi_{x_{0}}(M).$ Also $f(x_{0})^{-1}$ has been moved outside the integral.

            This expression shows that the wave packet expression is invariant under adding a constant to the fields, $\theta$ and $\phi.$  Note that changing the reference location to $y$ from $x_{0}$ simply replaces $x_{0}$ by $y$ in this expression.

            \subsubsection{Path lengths}

            Another example of  the effect of scaling is on the lengths of paths.  Let $q$ be a path on a space time manifold, $M,$ where $q$ is parameterized by real numbers $s.$ The initial and final points of $q$ are $q(0)=x$ and $q(1)=y.$  Let $\mathcal{RT}^{\cup}$ be the fiber bundle,\begin{equation}\label{MFRT}\mathcal{RT}^{\cup}=M\times\bigcup_{r}(\bar{R}^{r} \times\bar{T}^{r}\times\phi^{r}(M)),p,M.\end{equation}The fiber at $z$ is \begin{equation} \label{MFRTpz}p^{-1}(z)=\bigcup_{r}(\bar{R}^{r}_{z} \times\bar{T}^{r}_{z}\times\phi^{r}_{z}(M)).
            \end{equation}In these equations $\bar{T}^{r}_{z}$ is the local representation at $z$ of the tangent space at level $r$  on $M$.  Since $M$ is Minkowski, $\bar{T}^{r}_{z}$ covers all of $M$.  Here $\phi^{r}_{z}$ is a chart that maps $M$ onto a local level $r$ representation of $M$ at $z.$ In particular, $\phi^{r}_{z}(M)=(\textbf{R}^{r})^{4}_{z}.$

            As noted before,  the effects of number scaling on  theoretical descriptions of quantities is independent of the level at which they are described. An alternate way to say this is that the effects of number scaling on theory descriptions are invariant under level change.  For this reason the superscript $r$ will be suppressed by setting its value to be $r=1$. Thus $\textbf{R}^{4}_{z}$ is a local representation of $M$ at $z.$

            A local description in the fiber  at $z$ of the length of the path, $q,$ is given by \begin{equation}\label{Lqz}
            L(q)_{z}=\int_{z,0}^{1}|\nabla_{s}q_{z}\cdot\nabla_{s}q_{z}|^{1/2}ds_{z}.
            \end{equation}Here $q_{z}$ is the local representation of $q$ as a path on $\phi_{z}(M)$ where $q_{z}(s)=\phi_{z}(q(s)).$ Also $\nabla_{s}q_{z}$ is the gradient of $q_{z}$ at $s_{z}$ in the fiber at $z.$  No scaling is involved here because the path length is defined for a localized representation of $q$. Since $M$ is space time \begin{equation}\label{nbsqz}
             \nabla_{s}q_{z}\cdot\nabla_{s}q_{z}=\eta^{\mu,\mu}\partial_{\mu,s}q_{z}\partial_{\mu,s}q_{z}.
            \end{equation}Here $\eta^{\mu,\mu}$  for $\mu=0,1,2,3$ are the diagonal elements of the metric tensor for special relativity.

            The situation is quite different if one describes a global version of the length of $q$ and then localizes the description to obtain a meaningful result. The global version of the path length begins with the expression \begin{equation}\label{Lqq}L(q)_{g}=\int_{q,0}^{1} |\nabla_{s}q_{q(s)}\cdot\nabla_{s}q_{q(s)}|^{1/2}ds.\end{equation} In this expression $q_{q(s)}(s)$ is the point in $\textbf{R}^{4}_{q(s)}$ defined by $q_{q(s)}(s)=\phi_{q(s)}(q(s)).$ For each value of $s,$ the integrand is in the fiber at $q(s)$ on $M$.

            The integral in Eq. \ref{Lqq} is not defined because the integrands are in different fibers.  This is remedied by parallel transforming the integrands to a common fiber location, $x$, and then integrating. The result is given by \begin{equation}\label{Lqqx}L(q)_{g,x}=\int_{x,0}^{1} e^{-\theta(x)+\theta(q(s))_{x}}|\nabla_{s}q_{x}\cdot\nabla_{s}q_{x}|^{1/2}ds_{x}.\end{equation}Here $\theta(q(s)_{x})$ is the same real number in $\bar{C}^{f(x)}_{x}$ as $\theta(q(s))$ is in $\bar{C}^{f(q(s))}_{q(s)}.$

            The choice of reference point is arbitrary.  The equivalent expression for any other reference point, $z$ is obtained by multiplying $L(q)_{g,x}$ by $e^{-\theta(z)+\theta(x)}.$ The result is given by Eq. \ref{Lqqx} after replacing all occurrences of $x$ in the equation by $z.$

            The  global geodesic distance between $x$ and $y$ is affected by number scaling.   The resulting geodesic equation is obtained by variation of the path $q$ to obtain the Euler Lagrange equations.  Following the procedure in general relativity and replacing the variable $s$ by the proper time, $\tau$ gives the result \cite{Carroll,Ben1412}\begin{equation}\label{GExy}[\frac{d}{d\tau} +\textbf{$\Gamma$} \cdot\nabla_{\tau}q]\frac{dq^{\mu}}{d\tau} +\eta^{\mu,\mu}\Gamma_{\mu}(q(\tau))=0.
            \end{equation}Here, as before, \textbf{$\Gamma$} is the gradient of $\theta.$ There are four such equations, one for each value of $\mu.$

            The effect of scaling is shown by the presence of \textbf{$\Gamma$}.  If $\textbf{$\Gamma$}=0$,  the equation reduces to that for a straight line as  $d^{2}q^{\mu}/d\tau^{2}=0.$

            These are only two of the many examples of physical quantities that are affected by number scaling. Another effect is the replacement of the momentum operator by the canonical momentum.  This gives $\vec{p}\rightarrow \vec{p}+$\textbf{$\Gamma$}+i\textbf{$\Delta$}. Others include path integrals in quantum mechanics and the action as a space time integral of Lagrangian densities.

            These examples are sufficient to show the effect of number scaling on theory descriptions of  many physical quantities.  However experiments done to date do not show any effect of number scaling on physical quantities.  The results obtained here with scaling present  must be reconciled with this fact.

             \section{Theory and experiment}\label{TE}

            Reconciliation  begins with a  basic fact: \emph{All experiments, measurements, observations, and computations are done in a local region, $L$ of $M$.} Here $M$ represents background flat cosmological space and time. The region $L$ includes all locations occupied by us and occupiable, in principle, by other intelligent beings \emph{with whom we can establish effective two way communication.}

            The exact size of $L$ is not important. One does require that it be small compared to the volume of the universe.  One estimate in the literature sets $L$ to be a volume about $1200$ light years in radius presumably centered on the solar system \cite{Smith}.

            The nonobservance of the effect of the scaling field, $f$ in experiments means that the scalar fields $\theta$ and $\phi$ must be roughly constant over $L$.  Equivalently the corresponding gradients, \textbf{$\Gamma$} and \textbf{$\Delta$} must be too small to have shown up in  local experiments done to date.\footnote{Local experiments are those in which systems are prepared in some state and their properties measured, all within the region, $L$.} However this does not exclude the possibility that the scaling field  is cosmological, such as dark matter or dark energy.

            For locations outside $L$ there are no restrictions on these fields.  The reason is that there is no way to communicate with intelligent beings outside $L$ to determine if their observations, experiments and computations are consistent with ours where scaling effects are very small.  Such observers are just too far away.

            \subsection{Comparison of experiment and computation outcomes}\label{CECR}
            Each computation and experiment or measurement, as a physical process, necessarily occupies a finite region of space and time.  For purposes of discussion,  specific locations in $M$ will be associated with  computations and measurements.   In the presence of number scaling fields, one would think that scaling would affect the comparison of the output of a computation at point $x$ of $M$ with the output of an experiment or measurement at point $y.$

            If the output of the computation at $x$  is a number $r$ and the output of a measurement  at $y$ is a number $s$, then the number values associated with these numbers can be represented by $val_{f(x)}(r)$ and $val_{f(y)}(s).$ These are the numerical values or meanings associated with the computation and measurement outputs.

            Comparison of these values requires that they be compared locally at some common point. Let $y$ be the point.  Parallel transform of the computation output value, in the presence of the scaling field, $f,$ introduces a scaling factor. The result of the transform can be described by
            \begin{equation}\label{valyr}val_{f(y)}(r)=val_{f(y),f(x)}(val_{f(x)}(r))=\frac{f(y)}
            {f(x)}val_{f(x)}(r).\end{equation}Here $val_{f(y),f(x)}$ is the connection that relates number values in $\bar{R}^{f(x)}_{x}$ to number values in $\bar{R}^{f(y)}_{y}.$  The base sets for these structures are the same.

            This argument suggests that scaling affects comparisons between computation and measurement outputs. No hint of this is seen in physics.  This is explained by the fact that outputs of computations and measurements are not numbers.  They are physical systems in physical states. Comparison of outputs obtained at different locations requires physical transmission of the output states or the relevant information to a common location for  comparison and interpretation as number values.

            Interpretation of the output states as numbers in a base set, such as $R,$ followed by physical transmission of the numbers is reasonable.  In the presence of scaling,  physical transmission must occur before values or meanings are assigned to the outputs. Only if the ratio $f(y)/f(x)\simeq 1$ can parallel transform of number values be a valid comparison.

\begin{figure*}[h!]\begin{center}\vspace{3cm}
            \rotatebox{270}{\resizebox{200pt}{200pt}{\includegraphics[190pt,200pt]
            [540pt,550pt]{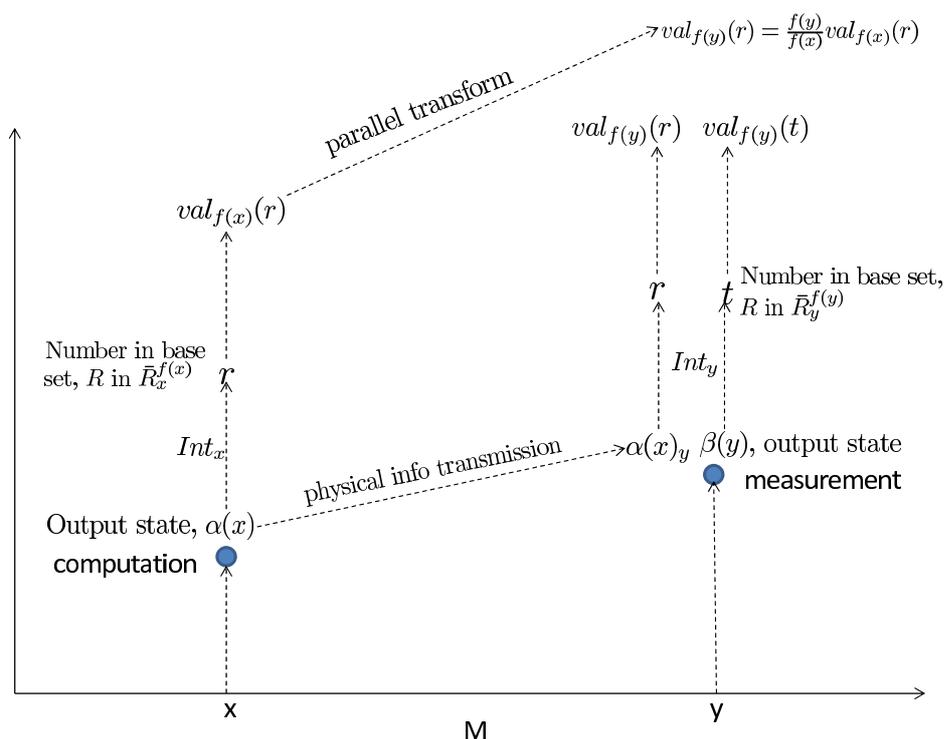}}}\end{center}
           \caption{Effect of number scaling on comparison of computation and experiment results is shown for two points $x$ and $y$ in $M$. Both physical transmission of outcome information and parallel transformation of the associated number values are shown.}\label{Ba.eps}\end{figure*}

            Figure \ref{Ba.eps} is a schematic representation of the effects of parallel transform of values and physical transmission of outcome state information as numbers.  It shows clearly the distinction between the two methods of comparison.
           The figure shows clearly a fundamental aspect of comparison of theory computations with experiment outcomes. It shows that the interpretation of the outcome states allows one to compare the outcomes as numbers and see if they agree or not.  The figure shows that the computation agrees with experiment if $r=t.$

            It is important to note that determination of the number value or meaning of these outcomes depends on $f$ or, equivalently, the values of $\theta$ and $\phi$ at the point of comparison. If the value of $f$ at the common location of comparison is not known, then the number values of the computation and experiment outcomes are not known.  They could have any value.

           This shows clearly that the scaling field $f$  mediates or interprets between  outcomes of physical processes, such as  measurements and computations, and theory.  Theoretical expressions   are all expressions based on values of mathematical quantities, such as scalars, vectors, operators, etc.  Predictions are usually expressed as numerical values of quantities.  As such these are number values in a number structure.

           As physical processes, experiments and computations yield numbers as outputs.  These numbers are elements of the base sets of number structures.  As such they can have any value.  The values of these numbers depend on $f$.  The value of any computation or measurement at point $x$ is determined by the value of $f$, or of $\theta$ and $\phi$, at $x.$  In essence $f$ acts as a connection or interpolation between theory and experiment.

           In this sense $f$ acts as as an essential link between theory and experiment.  The values of $f$ serve to assign meaning to the  numerical outcomes of computation or measurements.   If the values of $f$ are not known in the local regions in which the computations or experiments are implemented, then one cannot assign values to the outcomes.  The theory experiment and theory computation connections are broken.

            \section{Summary}\label{S}
            Number scaling of number structures in mathematics has been described with detailed examples given.  Scaling shows that two distinct concepts, that of number as elements of the base sets of structures and the values the numbers have in structure are conflated in the usual use of numbers. The presence of scaling shows that, by themselves, the elements of the base sets in number structures have no  intrinsic value.  they acquire values only as elements of the base set within structures. If $\bar{S}^{s}$  and $\bar{S}^{t}$ are scaled number structures with scaling parameters, $s$ and $t$, the value of a number $a$ in the base set $S$, is different in $\bar{S}^{s}$ than it is in $\bar{S}^{t}.$ The effect of number scaling for number structures extends of other types of mathematical structures, such as vector spaces, algebras, etc. that include scalars as part of their axiomatic description.

            Fiber bundles are used to describe the association of number structures and vector spaces to each point of a space and/or time manifold, $M$. The freedom of scaling results in the bundle fiber  consisting of a collection of all scaled number structures and vector spaces as in $\bigcup_{s}(\bar{S}^{s}\times\bar{V}^{s}).$ The properties of $M$  are such that the bundle over $M$  is a product bundle as in $M\times\bigcup_{s}(\bar{S}^{s}\times\bar{V}^{s}),p,M.$  The fiber at point $x$ of $M$ is given by  $p^{-1}(x)=\bigcup_{s}(\bar{S}^{s}_{x}\times\bar{V}^{s}_{x}).$

            A scalar scaling field, $f$, is introduced to describe the freedom of choice of scaling factor at each point of $M$.  The effect of this field on covariant derivatives of fields as sections on the fiber bundles shows up as the addition of  gradients of a pair of  scalar fields, $\theta(x)+i\phi(x),$  to the covariant derivatives in gauge theories and to any Lagrangian containing field derivatives. Here $f(x)=\exp(\theta(x)+i\phi(x)).$ Any mass, zero included, is possible for these fields. Possible physical candidates include the Higgs, the inflaton, dark matter, dark energy, or none of these.

            The effect of these fields on other nonlocal physical and geometric quantities, was shown for two examples, wave packets and path lengths.   These fields appear in the connections used  to parallel transform integrands in the  global expression for the wave packet to a reference location where the integration makes sense.

            The same concept applies to the localization of lengths of paths on $M$. The  real field, $\theta$, affects the distance between two points.  This is seen by the appearance of the gradient of $\theta$ in the geodesic equation.

            The lack of local physical evidence for  these fields imposes some restrictions.  One is that the gradients of both $\theta$ and $\phi$ must be close to $0$ in a local region of cosmological space and time. This is a region in which we, as observers, conduct experiments and computations.  It also includes regions within effective communication distance of us. This takes  account of the possibility that intelligent life exists outside the solar system.  There are no restrictions on the gradients of $\theta$ and $\phi$ outside the local region.

            Another relevant aspect of the presence of a scalar scaling field is that it has no effect on comparison of outcomes of computations or measurements with one another.  The reason is that these outcomes are physical systems in physical states.  Comparison requires transmission of the information in these states to a common location for local comparison. As a consequence the fact that the scaling field may have different values at the locations of the implementation of computations or experiments is not relevant.

            However the scaling field does affect the values assigned to the outcomes of computations or experiments.  These outcomes are numbers.  Values of these numbers are needed to compare with theoretical predictions as number values.  It follows that the field, $f$ serves as an interpolator  between process outcomes as numbers and values as theory predictions.  Comparison of theory values and experimental outcomes requires knowledge of the values of $f$ in the  local region. In the usual setup in the absence of scaling $f=1$ everywhere.

            \section{Outlook}\label{O}
            There is much to do in the future. Other nonlocal quantities such as momentum, actions and many other quantities are affected by scaling. These need to be investigated as well as other effects on the dynamics of quantum mechanical systems.  It might have some input into the measurement problem in quantum mechanics.

            In another direction, the manifold needs to be expanded to include general relativity. The effect of scaling on space and time needs work.  Finally one would hope to be able to link the scalar fields $\theta$ and $\phi$ directly with at least one of the scalar fields already described in the physics literature.

            \section*{Acknowledgement}
            This material is based upon work supported by the U.S. Department of Energy, Office of Science, Office of Nuclear Physics, under contract number DE-AC02-06CH11357.

            \end{document}